\documentclass[twoside]{dis07}
\usepackage[latin1]{inputenc}
\usepackage[dvips]{graphicx,epsfig,color}
\usepackage{wrapfig,rotating}
\usepackage{amssymb,amsmath,array}

\newcommand{\be}{\begin{equation}}
\newcommand{\ee}{\end{equation}}
\newcommand{\bea}{\begin{eqnarray}}
\newcommand{\eea}{\end{eqnarray}}
\newcommand{\bi}{\begin{itemize}}
\newcommand{\ei}{\end{itemize}}
\newcommand{\ben}{\begin{enumerate}}
\newcommand{\een}{\end{enumerate}}
\newcommand{\la}{\left\langle}
\newcommand{\ra}{\right\rangle}

\def\frac#1#2{{{#1}\over {#2}}}
\def\gsim{\mathrel{\rlap{\lower4pt\hbox{\hskip1pt$\sim$}}
    \raise1pt\hbox{$>$}}}         
\def\lsim{\mathrel{\rlap{\lower4pt\hbox{\hskip1pt$\sim$}}
    \raise1pt\hbox{$<$}}}         

\newcommand{\dat}{\mathrm{dat}}
\newcommand{\one}{\mathrm{(1)}}
\newcommand{\two}{\mathrm{(2)}}

\newcommand{\NS}{\mathrm{NS}}

\newcommand{\draft}[1]{}

\begin{document}
\title{Progress on Neural Parton Distributions}

\author{J.~Rojo$^1$, R.~D.~Ball$^2$, L. Del Debbio$^2$, S.~Forte$^3$, 
A.~Guffanti$^2$, \\ J.~I.~Latorre$^4$, A.~Piccione$^5$ and M.~Ubiali$^2$ 
(The NNPDF Collaboration)
%
%
\vspace{.3cm}\\
1.- LPTHE, CNRS UMR 7589,  Universit\'es Paris VI-Paris VII,
F-75252, Paris Cedex 05, France
\vspace{.1cm}\\
2.- School of Physics, University  of Edinburgh,
 Edinburgh EH9 3JZ, Scotland
\vspace{.1cm}\\
3.- Dipartimento di Fisica, Universit\`a di Milano, \\ and
INFN, Sezione di Milano, Via Celoria 16, I-20133 Milano, Italy
\vspace{.1cm}\\
4.- Departament d'Estructura i Constituents de la Mat\`eria, \\
Universitat de Barcelona, Diagonal 647, E-08028 Barcelona, Spain
\vspace{.1cm}\\
5.- INFN, Sezione di Genova, via Dodecaneso 33, I-16146 Genova,  Italy
}

\maketitle

\begin{abstract}
We give a status report on the determination of a set of parton
distributions based on neural networks. In particular, we 
summarize the determination
of the
nonsinglet quark distribution up to NNLO,
we compare it with results obtained using other approaches, and we
discuss its use for a determination of $\alpha_s$.
\end{abstract}

\section{Introduction}
The LHC will require
an approach to the search for
new physics based on the precision techniques which are
customary  at lepton machines~\cite{lh2,heralhc}. This has recently led to
significant progress in the determination of parton distribution
functions (PDFs) of the nucleon.
The main recent development
has been the availability of sets of PDFs with an estimate of the
associated uncertainty~\cite{Alekhin,CTEQ,MRST}. However, 
the standard approach to the  determination of the uncertainty on parton
distributions 
has several weaknesses, such as  the lack of control on
the bias due choice of a parametrization and, more in general, 
the difficulty in giving a consistent statistical interpretation to
the quoted uncertainties.

These problems have stimulated various new
approaches to the determination of PDFs~\cite{kosower}, in
particular the 
neural network approach, first proposed in Ref.~\cite{f2ns}. 
The basic idea  is to combine a
Monte Carlo
sampling of the probability measure on the space of
functions that one is trying to determine~\cite{kosower} with the use of neural
networks as universal unbiased interpolating functions.
In Refs.~\cite{f2ns,f2p} this strategy was successfully applied to a  somewhat
simpler problem, namely, the construction of a parametrization of
existing data on the DIS structure function $F_2(x,Q^2)$
of the  proton and neutron.
 The method was proven to be fast and
robust, to be amenable to detailed statistical studies, and to be in
many respects superior to conventional parametrizations of structure
functions based on a fixed functional form.

The determination of a parton set involves 
the significant complication of having to go from one or more
physical observables to a set of parton
distributions. 
Recently~\cite{nnqns} most of the
technical complications required for the construction of a neural
parton set have been tackled and solved in the process of constructing
a determination of the quark isotriplet parton distribution. This work
will be reviewed here. Also, based on this work, we will present
preliminary results on the  determination of $\alpha_s$ and a
determination  of the 
variation in $\chi^2$ which corresponds to a one-sigma variation of
the underlying parton distributions.
Work to apply the techniques of~\cite{nnqns} to the singlet sector
is at an advanced stage~\cite{nnsinglet}.

\section{Determination of the  nonsinglet quark distribution}
The first 
application of the neural network approach to
parton distributions, a determination of the
NS parton distribution $q_{\NS}(x,Q_0^2)=(u+\bar{u}-d+\bar{d})(x,Q_0^2)$
from the DIS structure function data of the NMC
and BCDMS collaborations, was presented in Ref.~\cite{nnqns}. 
Results for this PDF were obtained at LO, NLO, NNLO
for different values of $\alpha_s(M_Z^2)$. 

In Ref.~\cite{nnqns} we have implemented a 
new fast and efficient method for solving 
the evolution equations up to NNLO. This
method 
combines the advantages
of $x-$space and $N-$space evolution codes: an $x$ dependent Green
function (evolution factor) is determined by  inverse Mellin transformation
of the exact $N$-space expression and stored. Evolution of PDFs is
then  performed by
convoluting this Green function with any given boundary condition.
The accuracy of this method has been benchmarked up to NNLO with the
help of the tables of Refs.~\cite{lh2,heralhc}.

Also, we have implemented a criterion to determine the convergence of
the fitting procedure in a way which is free of bias
related to the choice of parametrization. To this purpose, 
the
dataset  is randomly divided into two sets, of which only one is used
in the fit. Convergence is achieved when the quality of the fit to
data which are {\it not} used for minimiztion stops improving.

An important feature of our approach is that it is possible to 
check quantitatively the statistical features of  results using
suitable estimators. For example, one can check that the results  do not
depend on choices  made during the fitting procedure, such as
 the choice of architecture of neural networks, which is analogous to
the choice of parton parametrization in conventional fits. Namely, we 
repeat the fit with a different choice, and compute the distance
\be
\label{dist}
d[q]=\sqrt{\la\frac{\left( q_i^\one-
    q_i^\two\right)^2} {(\sigma_i^\one)^2+(\sigma_i^\two)^2}\ra_\dat
},
\ee
where $ q_i^\one$, $q_i^\two$ 
are  the predictions for the $i$-th data
point in the two fits, and $\sigma_i^\one$, $\sigma_i^\two$ the
predictions for the corresponding statistical uncertainties, and the average is
performed over all data. The results of the first and second fit are
the same if  $ d[q]= 1$ on average. This also checks that the
statistical uncertainties are correctly estimated. One can similarly
check stability of the uncertainty estimate.
In Ref.~\cite{nnqns} this comparison has been performed succesfully.

\begin{figure}
\centering
\includegraphics[width=0.77\columnwidth]{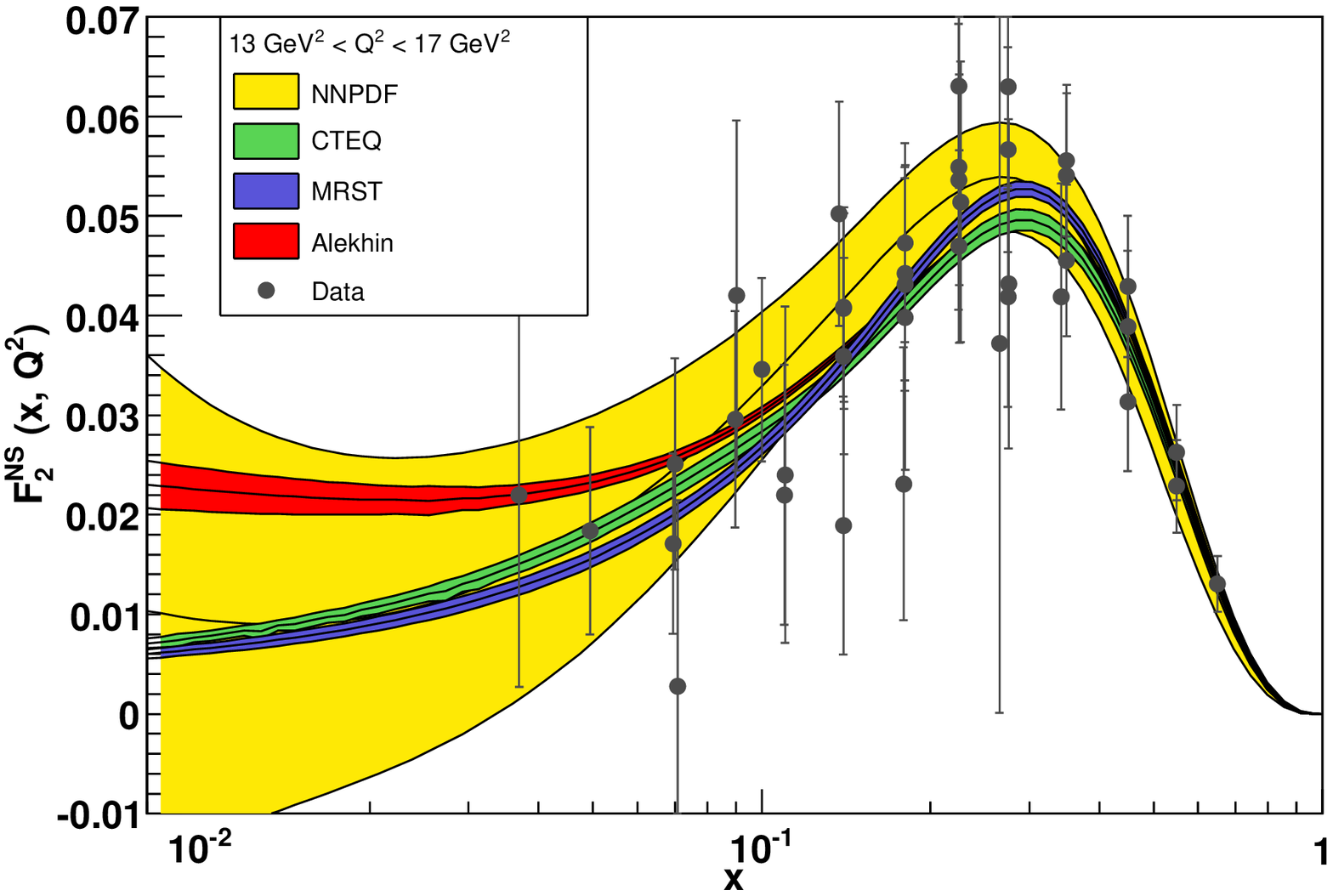}
\caption{}{The nonsinglet 
 structure function  $F_2^{\rm NS}$ as determined by the NNPDF
 collaboration~\cite{nnqns} from 229 NMC and 254 BCDMS data points, compared 
  to data  and various other determinations.}\label{Fig:f2nsplot}
\end{figure}

In Fig.~\ref{Fig:f2nsplot} we compare our results for the
NS structure function $F_2^{\NS}$ to other published
determinations. These 
results are available through the
webpage of the NNPDF Collaboration: 
 {\tt http://sophia.ecm.ub.es/nnpdf}. 
The large uncertainty that we find is a genuine feature of the
determination  of the nonsinglet quark distribution from the data
included in our fit,
and, especially at small $x$, it  appears to reflect the current
knowledge of the nonsinglet quark distribution.
Indeed,  for $x \le 0.05$
the only data which constrain the $q_{\NS}$ combination in global fits are the 
data used in the determination of Ref.~\cite{nnqns}. Hence, our results suggest
that standard fits might
be underestimating PDF uncertainties. 

In recent work on PDF uncertainties~\cite{CTEQ,MRST} it has been
suggested that, mostly because of inconsistencies between data, the
variation of the total $\chi^2$ which corresponds to a one--sigma
variation of the underlying PDFs is of order of $\Delta\chi^2\sim 50$
for the global fits presented in those references instead of $\Delta\chi^2=1$
of a statistically consistent fit~\cite{Alekhin}. In our approach,
this quantity can be computed. We get 
$\Delta \chi^2 \approx 1.7$ (preliminary). This implies that the NMC
and BCDMS data are mostly consistent, though some inconsistent data
are present~\cite{f2ns,f2p}. An extensive discussion of the way the
published~\cite{nnqns} and forthcoming ~\cite{nnsinglet} fits based on
the neural network approach can be used for the determination of
physical parameters (such as $\alpha_s$) and statistical properties of
the data (such as $\Delta \chi^2$) will be presented in a forthcoming
publication. 

\begin{figure}
\centering
\includegraphics[width=0.77\columnwidth]{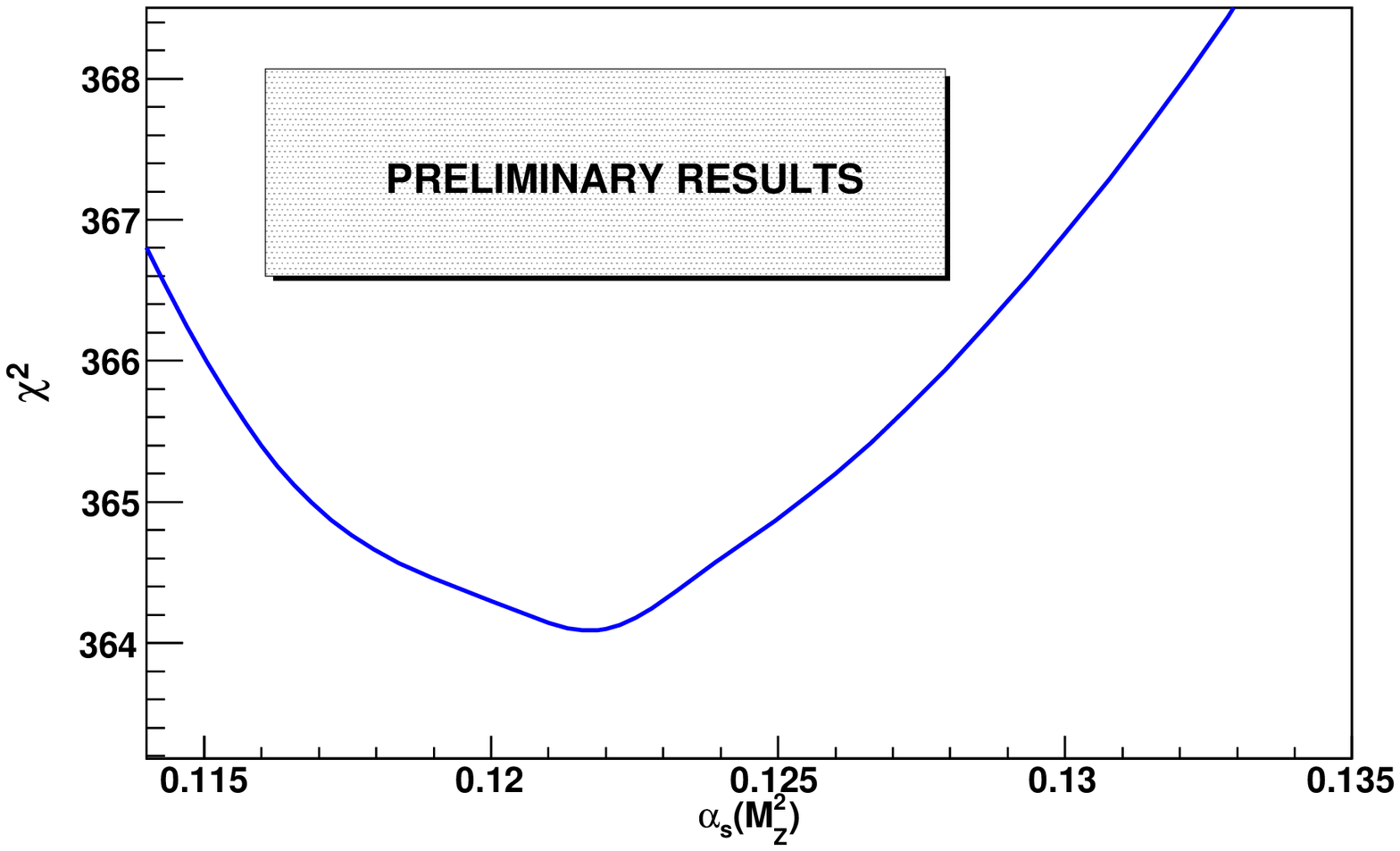}
\caption{}{The $\chi^2$ profile for a preliminary NNLO
determination of $\alpha_s(M_Z^2)$ from NS data. The number 
of data points included in the fit is $N_{\dat}=483$.}\label{Fig:alphafitplot}
\end{figure}

In~\cite{nnqns} the strong coupling $\alpha_s(M_Z^2)$ was 
fixed, but we could also extract it from the fit.
The results of 
a preliminary analysis, shown in Fig.~\ref{Fig:alphafitplot}, suggest
that nonsinglet data determine $\alpha_s(M_Z^2)$ with an uncertainty
which is  rather larger  than that ($\Delta\alpha_s(M_Z^2)\sim 0.002$)
obtained in comparable determinations (see e.g.
Ref.~\cite{blum}). This preliminary result is consistent with that obtained 
using the same data in Ref.~\cite{trun}, with a method
which eliminates the need to choose a parton parametrization. 
This strengthens the conclusion
that uncertainties in available PDF fits might be underestimated.

\section{Towards a full parton set}
The extension of the results described in Ref.~\cite{nnqns} to a full
global PDF fit has benefited from the increased manpower of the NNPDF
Collaboration, and is at a rather advanced stage~\cite{nnsinglet}.
In particular, the evolution formalism of Ref.~\cite{nnqns} has been
extended to the computation of a full set of neutral-current and charged-current
structure functions and fully benchmarked. A first full neural parton fit
is in is in preparation. It will at first be
based on DIS data only, including all  available $F_2^p$ and
$F_2^d$ fixed target data and  the full NC and CC
HERA reduced cross sections. 


\begin{footnotesize}


\end{footnotesize}


\end{document}